\documentstyle[12pt]{article}

\newcommand{\sect}[1]{\setcounter{equation}{0}\section{#1}}

\newcommand{\bea}{\begin{eqnarray}}
\newcommand{\eea}{\end{eqnarray}}
\newcommand{\be}{\begin{equation}}
\newcommand{\ee}{\end{equation}}
\newcommand{\vs}[1]{\vspace{#1 mm}}

\newcommand{\dsl}{\pa \kern-0.5em /}

\newcommand{\pa}{\partial}

\newcommand{\nn}{\nonumber\\}

\begin{document}
\topmargin 0pt
\oddsidemargin 0mm

%\renewcommand{\thefootnote}{\fnsymbol{footnote}}
%\begin{titlepage}
\begin{flushright}
USTC-ICTS-04-20\\
MCTP-04-52\\
hep-th/0409019\\
%SINP-TNP/02-7
\end{flushright}

\vs{2}
\begin{center}
{\Large \bf
Delocalized, non-SUSY $p$-branes, tachyon condensation
and tachyon matter}
\vs{10}

{\large J. X. Lu$^a$\footnote{E-mail: jxlu@ustc.edu.cn}
 and Shibaji Roy$^b$\footnote{E-mail: roy@theory.saha.ernet.in}}

 \vspace{5mm}

{\em
 $^a$ Interdisciplinary Center for Theoretical Study\\
 University of Science and Technology of China, Hefei, Anhui 230026,
P. R. China\\
and\\
Interdisciplinary Center of Theoretical Studies\\
Chinese Academy of Sciences, Beijing 100080, China\\
 and\\

Michigan Center for Theoretical Physics\\
Randall Laboratory, Department of Physics\\
University of Michigan, Ann Arbor, MI 48109-1120, USA\\

\vs{4}

 $^b$ Saha Institute of Nuclear Physics,
 1/AF Bidhannagar, Calcutta-700 064, India}
\end{center}

\vs{5}
\centerline{{\bf{Abstract}}}
\vs{5}
We construct non-supersymmetric $p$-brane solutions of type II supergravities
in arbitrary dimensions ($d$) delocalized in one of the spatial transverse
directions. By a Wick rotation we convert these solutions into Euclidean
$p$-branes delocalized in the transverse time-like direction. The former
solutions in $d=10$ nicely interpolate between the $(p+1)$-dimensional
non-BPS D-branes and the $p$-dimensional BPS D-branes very similar to 
the picture
of tachyon condensation for the tachyonic kink solution on the non-BPS 
D-branes. On the other hand the latter solutions interpolate between the 
$(p+1)$-dimensional non-BPS D-branes and the tachyon matter supergravity
configuration very similar to the picture of rolling tachyon on the non-BPS
D-branes.
\newpage

\section{Introduction}

In \cite{lrone} we constructed static, non-supersymmetric $p$-brane solutions
of type II supergravities in $d$-dimensions and showed how they interpolate
between chargeless $p$-brane--anti $p$-brane solutions and BPS $p$-brane 
solutions. So, the two different brane solutions of the two sides of the
interpolation have the same dimensionalities. However, for the case of 
non-BPS branes, the tachyon condensation on the kink solution reduces the
non-BPS D($p+1$)-branes to codimension one BPS D$p$-branes \cite{asone,astwo}. 
Therefore, the
two brane solutions in this case have dimensionalities differing by one.
 The most natural 
way to see this picture emerging from a supergravity solution (in the absence
of explicit appearance of the tachyon field) is to consider the 
non-supersymmetric $p$-brane solutions delocalized in one of the transverse 
spatial directions. The purpose of this paper is to construct such solutions
and study their properties, in particular, we will try to understand how
the BPS D$p$-branes arise from the non-BPS D$(p+1)$-branes \cite{astwo}
and also how the
supergravity configuration of tachyon matter \cite{asthree,asfour,oy}
arise from these solutions.

For BPS D-branes, the difference in dimensionalities (i.e. D$p$ $\to$
D$(p+1)$, or D$(p+1)$ $\to$ D$p$) appear due to T-duality transformation
and in this process the theory also changes from type IIA (IIB) to type
IIB (IIA). For example, to construct a D$(p+1)$-brane from a D$p$-brane,
one first delocalizes the D$p$-brane solution in type IIA (or IIB) theory
by placing a continuous array of D$p$-branes along one of the transverse
spatial directions (the T-dual direction). This produces an isometry in that
particular direction and then the application of T-duality along this
direction produces a localized D$(p+1)$-brane solution in type IIB (or IIA)
theory \cite{lps}. This procedure works because the BPS branes 
do not interact with each
other. However, because the non-supersymmetric branes interact, it is not 
clear how the above process of delocalization will work. This is the reason
we have to explicitly solve the equations of motion of type II supergravities
containing a metric, a dilaton and a $q=d-p-2$ form field-strength. We use
a specific ansatz for the metric and the form-field to solve the equations
of motion and obtain delocalized, non-supersymmetric $p$-branes characterized
by four independent parameters. We show that, unlike the BPS $p$-branes,
it is possible to convert these solutions to fully localized $(p+1)$-branes,
without taking T-duality, if the parameters satisfy certain condition. We
recognize these to be the non-BPS D$(p+1)$-brane solutions 
\cite{bmo,divec,lrone} of the same 
theory as the original $p$-brane solutions. This also explains why we have
non-BPS D-branes of odd and even dimensionalities in type IIA and type 
IIB string theories respectively \cite{asone}. By scaling certain 
parameters of the
delocalized solutions appropriately, we show how these solutions reduce to 
BPS D$p$-brane solutions. We therefore interpret these solutions as the
interpolating solutions between non-BPS D$(p+1)$-branes and the BPS 
D$p$-branes very similar to the tachyon condensation for the tachyonic kink
solution on the non-BPS D$(p+1)$-branes \cite{asone,astwo}.

Next we Wick rotate these solutions which simply exchanges the delocalized
spatial direction with the time-like direction of the non-supersymmetric
$p$-branes. We therefore end up getting Euclidean $p$-branes delocalized
along the transverse time-like direction of the branes. We find that these 
solutions will be real only if they do not possess any charge and so, they
are characterized by three parameters. As discussed before, these
solutions can also be converted to fully localized $(p+1)$-branes i.e.
non-BPS D$(p+1)$-branes if the parameters satisfy certain condition. However,
since in this case one can not have charged solutions, we find that it is
not possible to obtain completely localized Euclidean $p$-branes (or S-branes)
\cite{stro,cgg,kmp,roy,ohta}
from the delocalized solutions by scaling parameters as was mentioned for
the spatially delocalized  case. On the other hand, we show that 
by adjusting the parameters
it is possible to obtain the supergravity configuration of tachyon matter
\cite{oy} from these Wick rotated solutions. We therefore interpret 
these solutions
as the interpolating solutions between non-BPS D$(p+1)$-branes and the 
tachyon matter very similar to the rolling tachyon solution \cite{asfive}
of the non-BPS 
D$(p+1)$-branes discussed by Sen.

This paper is organized as follows. In section 2, we construct and discuss
the properties of spatially delocalized, non-symmetric $p$-branes. The Wick 
rotated versions and their properties are discussed in section 3. We conclude
in section 4.

\sect{Spatially delocalized, non-SUSY $p$-branes}

In this section we construct and study the properties of the 
non-supersymmetric $p$-branes in $d$-dimensions delocalized along
one of the $(d-p-1)$ spatial transverse directions. The relevant supergravity
action in the Einstein frame has the form,
\be
S = \int d^dx \sqrt{-g} \left[R - \frac{1}{2} \partial_\mu \phi \partial^\mu
\phi - \frac{1}{2\cdot q!} e^{a\phi} F_{[q]}^2\right]
\ee
where $g_{\mu\nu}$, with $\mu,\,\nu = 0,1,\ldots,d-1$ is the metric and
$g={\rm det}(g_{\mu\nu})$, $R$ is the scalar curvature, $\phi$ is the dilaton,
$F_{[q]}$ is the field strength of a $(q-1) = (d-p-3)$-form gauge field
and $a$ is the dilaton coupling.

We will solve the equations of motion following from (2.1) with the ansatz
for the metric and the $q$-form field strength as given below,
\bea
ds^2 &=& e^{2A(r)}\left(dr^2 + r^2 d\Omega_{d-p-3}^2\right) + e^{2B(r)} \left(
-dt^2 + dx_1^2 + \cdots + dx_{p}^2\right) + e^{2C(r)} dx_{p+1}^2\nn
F_{[q]} &=& b\,\, {\rm Vol}(\Omega_{d-p-3}) \wedge dx_{p+1}
\eea
In the above $r = (x_{p+2}^2 + \cdots + x_{d-1}^2)^{1/2}$, $d\Omega_{d-p-3}^2$
is the line element of a unit $(d-p-3)$-dimensional sphere,
Vol($\Omega_{d-p-3}$) is its volume-form and $b$ is the magnetic charge
parameter. The solutions (2.2) represent magnetically charged $p$-branes
delocalized in a transverse spatial direction $x_{p+1}$. It should be clear
from the form of $r$ given before which says that the true transverse
directions are $x_{p+2}, \ldots, x_{d-1}$. On the other hand $x_{p+1}$
is neither a transverse direction nor a brane direction (since $B(r) \neq
C(r)$ in general), but represents the delocalized direction. We will also
use a gauge condition
\be
(p+1) B(r) + (q-2) A(r) + C(r) = \ln G(r)
\ee
Note that when $G(r)=1$, the above condition reduces to the extremality 
or supersymmetry condition \cite{dkl}. We call $G(r)$ as the 
non-extremality function,
whose extremal limit is $G(r) \to 1$.

Using (2.2) and (2.3) the various components of Einstein equation and the 
dilaton equation take the forms, 
\bea
B'' + \frac{q-1}{r}B' + \frac{G'}{G} B' - \frac{b^2(q-1)}{2(d-2)}
\frac{e^{2(p+1)B+a\phi}} {G^2 r^{2(q-1)}} &=& 0 \\
C'' + \frac{q-1}{r}C' + \frac{G'}{G} C' + \frac{b^2(p+1)}{2(d-2)}
\frac{e^{2(p+1)B+a\phi}} {G^2 r^{2(q-1)}} &=& 0\qquad\qquad \\
A'' + \frac{q-1}{r}A' + \frac{G'}{G} \left(A' + \frac{1}{r}\right) 
+ \frac{b^2(p+1)}{2(d-2)}
\frac{e^{2(p+1)B+a\phi}} {G^2 r^{2(q-1)}} &=& 0\qquad\qquad \\
-A'' - \frac{G''}{G} + \frac{G'^2}{G^2} - \frac{1}{p+1}\left(\frac{G'}{G}
- (q-2)A' - C'\right)^2 - (q-2) A'^2 & &\nn
+ \frac{G'}{G} A' - \frac{q-1}{r} A' 
- C'^2 - \frac{1}{2} {\phi'}^2
+ \frac{b^2(q-1)}{2(d-2)} \frac{e^{2(p+1)B + a\phi}}{G^2 r^{2(q-1)}} &=& 0
\qquad\qquad\\
{\phi''} + \frac{q-1}{r}{\phi'} + \frac{G'}{G}{\phi'} -
\frac{a b^2}{2}\frac{e^{2(p+1)B+a\phi}}{G^2 r^{2(q-1)}} &=& 0\qquad\qquad
\eea
In the above `prime' denotes derivative with respect to $r$. Using (2.3),
(2.4) and (2.5) in eq.(2.6) we obtain an equation for the non-extremality
function as,
\be
\frac{G''}{G} + \frac{(2q-3)}{r} \frac{G'}{G} = 0
\ee
There are three different solutions to this equation and they are
\be
(i)\,\,\, G = 1 - \frac{\omega^{2(q-2)}}{r^{2(q-2)}},
\qquad (ii)\,\,\, G = 1 + \frac{\omega^{2(q-2)}}{r^{2(q-2)}},
\qquad (iii)\,\,\, G = \frac{\omega^{2(q-2)}}{r^{2(q-2)}}
\ee
The solution in $(iii)$ can be shown to be supersymmetric by a coordinate
transformation and lead to the near horizon limits of delocalized BPS
$p$-brane solutions \cite{lrone}. Since we are interested in 
non-supersymmetric solutions
we do not consider $(iii)$. Also the solution $(ii)$ is not of our
interest since it gives  non-supersymmetric delocalized $p$-brane
solutions which have BPS limits leading to some unusual brane configuration
and not the usual BPS $p$-brane configuration \cite{lrone}.
Since we are interested in interpolating solutions between non-BPS 
D$(p+1)$-branes and the usual BPS D$p$-branes we will consider only 
case $(i)$. The
non-extremality function in this case can be factorized as follows,
\be
G(r) = \left(1+\frac{\omega^{q-2}}
{r^{q-2}}\right)\left(1-\frac{\omega^{q-2}}{r^{q-2}}\right) =
H (r) {\tilde H} (r)
\ee
where $H(r) = 1 + \omega^{q-2}/r^{q-2}$, ${\tilde H}(r) = 
1 - \omega^{q-2}/
r^{q-2}$, with $\omega^{q-2}$, a real parameter. Now from (2.4) and (2.8) 
we express $\phi$ in terms of $B$ and also from (2.4) and (2.5) we express $C$
in terms of $B$ as follows,
\bea
\phi &=& \frac{a(d-2)}{q-1} B + \delta_1 \ln \frac{H}{\tilde{H}}\\
C &=& -\frac{p+1}{q-1}B + \delta_2 \ln \frac{H}{\tilde{H}}
\eea
where $\delta_1$ and $\delta_2$ are two real and negative integration
constants which can be understood if one actually finds the above
solutions from the corresponding equations of motion. We can 
also determine $A$ in terms of $B$ using (2.3) and (2.13) as, 
\be
A = -\frac{p+1}{q-1}B - \frac{\delta_2}{q-2} \ln \frac{H}{\tilde{H}}
+ \frac{1}{q-2} \ln (H\tilde{H})
\ee
We therefore have to solve $B$ from eq.(2.4) and check whether the
solution is consistent with eq.(2.7). In order to solve $B$ we make an ansatz
\bea
e^B &=& F^{\gamma}\nn
{\rm with},\qquad F &=& \cosh^2\theta \left(\frac{H}{\tilde 
{H}}\right)^\alpha - \sinh^2\theta
\left(\frac{\tilde {H}}{H}\right)^\beta
\eea
where $\alpha$, $\beta$, $\theta$ and $\gamma$ are real constants and we will
comment on them later. Substituting (2.15) in eq.(2.4) we find that the
solutions exist provided the parameters satisfy the following relations,
\be
\gamma \chi = -2, \quad \alpha-\beta \,\,=\,\, a\delta_1, \quad
b = \sqrt{\frac{4 (d-2)}{\chi (q-1)}} (q-2)(\alpha+\beta) \omega^{q-2}
\sinh 2\theta
\ee
where $\chi = 2(p+1) + a^2(d-2)/(q-1)$. Note that we have taken $b \,
\geq 0$ and kept both sign choices for   
$\alpha+\beta$ for later convenience. The sign of 
$\alpha+\beta$ determines the sign for $\theta$, given 
$b > 0$ in (2.16). These solutions
are consistent with eq.(2.7) provided the parameters satisfy
\be
\frac{1}{2} \delta_1^2 + \frac{2\alpha(\alpha-a\delta_1)(d-2)}{\chi(q-1)}
= (1-\delta_2^2)\frac{q-1}{q-2}
\ee
From (2.17) and $\alpha-\beta = a \delta_1$, we can express $\alpha$ and
$\beta$ in terms of $\delta_1$ and $\delta_2$ as,
\bea
\alpha &=& \pm \sqrt{\frac{\chi (q-1)^2
}{2(d-2)(q-2)} (1-\delta_2^2) - \frac{\delta_1^2}{2}\frac{(q-1)(p+1)}
{(d-2)}}
+ \frac{a\delta_1}{2}\nn
\beta &=& \pm \sqrt{\frac{\chi (q-1)^2
}{2(d-2)(q-2)} (1-\delta_2^2) - \frac{\delta_1^2}{2}\frac{(q-1)(p+1)}
{(d-2)}}
- \frac{a\delta_1}{2}
\eea
We thus find from the above relations that both $\delta_2$ and $\delta_1$ are
bounded by\footnote{The solutions also exist when these bounds are
violated \cite{lrone} but their BPS limits do not give the usual BPS 
$p$-branes, 
therefore they, as mentioned
earlier, are not considered in this paper.}  
\bea
|\delta_2| &\leq & 1\nn
|\delta_1| &\leq & \sqrt{\frac{\chi (q-1) (1-\delta_2^2)}{(q-2)(p+1)}}
\eea
Since we found $\gamma = -2/\chi$, we obtain from (2.12) -- (2.15)
\bea
e^{2A} &=& F^{\frac{4(p+1)}{(q-1)\chi}}
(H {\tilde {H}})^{\frac{2}{q-2}} \left(\frac{H}{\tilde{H}}\right)^{-\frac{
2\delta_2}{q-2}}\nn
e^{2B} &=& F^{-\frac{4}{\chi}}\nn
e^{2C} &=& F^{\frac{4(p+1)}{(q-1)\chi}}
\left(\frac{H}{\tilde{H}}\right)^{2\delta_2}\nn
e^{2\phi} &=& F^{-\frac{4a(d-2)}{(q-1)\chi}}
\left(\frac{H}{\tilde{H}}\right)^{2\delta_1}
\eea
So, the complete non-supersymmetric $p$-brane solutions delocalized in 
transverse $x_{p+1}$ direction have the forms,
\bea
ds^2 &=& F^{\frac{4(p+1)}{(q-1)\chi}} (H{\tilde {H}})^{\frac{2}{q-2}}\left(
\frac{H}{\tilde{H}}\right)^{-\frac{2\delta_2}{q-2}}
\left(dr^2 + r^2
d\Omega_{d-p-3}^2\right) + F^{-\frac{4}{\chi}}(-dt^2 + \sum_{i=1}^p
dx_i^2)\nn
& & \qquad\qquad\qquad\qquad\qquad\qquad\qquad\qquad\qquad + 
F^{\frac{4(p+1)}{(q-1)\chi}}\left(
\frac{H}{\tilde{H}}\right)^{2\delta_2} dx_{p+1}^2\nn
e^{2\phi} &=& F^{-\frac{4a(d-2)}{(q-1)\chi}}
\left(\frac{H}{\tilde {H}}\right)^{2\delta_1},\qquad
F_{[q]}\,\, =\,\, b\,\,{\rm Vol}(\Omega_{d-p-3}) \wedge dx_{p+1}
\eea
Note that unlike the localized solutions \cite{zz,lrone}, which are 
characterized by three
parameters, the delocalized solutions are charaterized by four parameters
$\omega$, $\theta$, $\delta_1$ and $\delta_2$ ($\alpha$ and $\beta$ are given
in terms of $\delta_1$ and $\delta_2$ as in (2.18) and $b$ is related to
$\delta_1$, $\delta_2$, $\omega$ and $\theta$ by (2.16)). We thus find
that the delocalization actually introduces one more parameter in the
non-supersymmetric solutions and this does not happen for BPS solutions. This
will prove crucial to interpret these solutions as interpolating solutions
between non-BPS D$(p+1)$ branes and BPS D$p$-branes. Also since these
solutions are non-supersymmetric, the four parameters would presumably be
related to the mass, the charge, the tachyon vev $\langle T \rangle$
and the vev of its derivative $\langle \partial_x T\rangle$ of the
non-supersymmetric $p$-branes. However, the microscopic string interpretation
of these solutions and also the precise relationships of these parameters
and the physical parameters just mentioned are not clear to us. 

In $d=10$, the delocalized $p$-brane solutions (2.21) take the forms,
\bea
ds^2 &=& F^{\frac{p+1}{8}} (H{\tilde {H}})^{\frac{2}{6-p}}\left(
\frac{H}{\tilde{H}}\right)^{-\frac{2\delta_2}{6-p}}
\left(dr^2 + r^2
d\Omega_{7-p}^2\right) + F^{-\frac{7-p}{8}}(-dt^2 + \sum_{i=1}^p
dx_i^2)\nn
& & \qquad\qquad\qquad\qquad\qquad\qquad\qquad\qquad\qquad + 
F^{\frac{p+1}{8}}\left(
\frac{H}{\tilde{H}}\right)^{2\delta_2} dx_{p+1}^2\nn
e^{2\phi} &=& F^{-a}
\left(\frac{H}{\tilde {H}}\right)^{2\delta_1},\qquad
F_{[q]}\,\, =\,\, b\,\,{\rm Vol}(\Omega_{7-p}) \wedge dx_{p+1}
\eea
where $F$ is as given in (2.15) and $H=1+\omega^{6-p}/r^{6-p}$,
$\tilde{H} = 1-\omega^{6-p}/r^{6-p}$ and also $\chi=32/(7-p)$.

Once we know the form of the metric, we can calculate the energy-momentum
(e-m) tensor associated with the brane from the linearized form of Einstein
equation given by,
\be
\nabla^2 \left(h_{\mu\nu} -\frac{1}{2} \eta_{\mu\nu}
h\right) = -2 \kappa_0^2 T_{\mu\nu} \delta^{(8-p)}(r)
\ee
where we have expanded the metric around asymptotically flat space as
$g_{\mu\nu} = \eta_{\mu\nu} + h_{\mu\nu}$ and used the harmonic gauge
$\partial_\lambda h^\lambda_\mu - \frac{1}{2} \partial_\mu h = 0$ with
$h = \eta^{\mu\nu} h_{\mu\nu}$.  Also in (2.23) $2\kappa_0^2=16\pi G_{10}$,
$G_{10}$ being the ten dimensional Newton's constant. From (2.22) we find
\bea
h_{00} &=& \frac{7-p}{8}\left[(\alpha+\beta)\cosh2\theta+(\alpha-\beta)\right]
\frac{\omega^{6-p}}{r^{6-p}}\nn
h_{ij} &=& -\frac{7-p}{8}\left[(\alpha+\beta)\cosh2\theta+(\alpha-\beta)\right]
\frac{\omega^{6-p}}{r^{6-p}}\delta_{ij}\nn
h_{xx} &=& \left\{\frac{p+1}{8}\left[(\alpha+\beta)\cosh2\theta+(\alpha-\beta)
\right] + 4 \delta_2 \right\}
\frac{\omega^{6-p}}{r^{6-p}}\nn
h_{mn} &=& \left\{\frac{p+1}{8}\left[(\alpha+\beta)\cosh2\theta+(\alpha-\beta)
\right] - \frac{4\delta_2}{6-p}\right\}
\frac{\omega^{6-p}}{r^{6-p}}\delta_{mn}\nn
h &=& \left\{\frac{p+1}{4}\left[(\alpha+\beta)\cosh2\theta+(\alpha-\beta)
\right] - \frac{8\delta_2}{6-p}\right\}
\frac{\omega^{6-p}}{r^{6-p}}
\eea
where $i,j=1,\ldots,p$, $x=x_{p+1}$ and $m,n=p+2,\ldots,9$. Substituting (2.24)
in (2.23) we obtain,
\bea
T_{00} &=& \frac{\Omega_{7-p}}{2\kappa_0^2}(6-p)\omega^{6-p}
\left[(\alpha+\beta)\cosh2\theta+(\alpha-\beta) - 
\frac{4\delta_2}{6-p}\right]\nn
T_{ij} &=& -\frac{\Omega_{7-p}}{2\kappa_0^2}(6-p)\omega^{6-p}
\left[(\alpha+\beta)\cosh2\theta+(\alpha-\beta) - 
\frac{4\delta_2}{6-p}\right]\delta_{ij}\nn
T_{xx} &=& \frac{\Omega_{7-p}}{2\kappa_0^2}(6-p)\omega^{6-p}
\left[ 
\frac{4\delta_2 (7-p)}{6-p}\right]\nn
T_{mm} &=& 0
\eea
Here $\Omega_n = 2 \pi^{(n+1)/2}/\Gamma((n+1)/2)$ is the volume of the
$n$-dimensional unit sphere.
In the above $T_{00}$ is nothing but the ADM mass of the brane. It has the
dimensionality mass per unit $(p+1)$-brane volume and therefore shows that
the energy is spread also along the delocalized direction $x=x_{p+1}$ as
expected. The fact that the brane is spread along $x$ can also be seen 
from $T_{xx}$ in (2.25) which is 
non-vanishing. $T_{mm}=0$ implies that the brane is localized along 
$m=x_{p+2},\ldots, x_{d-1}$ directions and they are the true transverse 
directions.

Now let us look at the metric in (2.22). These represent non-supersymmetric 
$p$-branes delocalized in $x_{p+1}$ direction in $d=10$. Note that for
BPS case one can make such solutions localized $(p+1)$-brane by a T-duality
transformation and so if the $p$-brane is a solution to type IIA (or IIB)
theory then $(p+1)$-brane is a solution of type IIB (or IIA) theory. However,
in this case it is possible to make the $p$-brane solution to a localized
$(p+1)$-brane without taking T-duality by simply putting $\theta=0$ and
$2\delta_2 = -\alpha$ (note that this is possible because of the presence
of the extra parameters which are not present for the BPS
solutions. Here we choose a plus sign in (2.18).). To 
make it clear note from the last two terms of the metric in (2.22) that for 
the coefficients of these two terms to match (which is necessary to make
it a metric for localized $(p+1)$-brane) $F$ must be some powers of 
$(H/\tilde{H})$ and from (2.15) we find that this happens only for $\theta
=0$. The coefficients would then match for $\alpha=-2\delta_2$. From (2.16)
we see that $\theta=0$ corresponds to $b=0$ i.e. the solutions must be 
chargeless. Also from the expressions of the e-m tensors we see that for
$\theta=0$ and $\alpha=-2\delta_2$, $T_{00} = - T_{ii}$ for $i=1,\ldots,
(p+1)$, i.e. we have a localized $(p+1)$-brane. The solutions and the e-m
tensors then take the forms,
\bea
ds^2 &=& (H{\tilde {H}})^{\frac{2}{6-p}}\left(
\frac{H}{\tilde{H}}\right)^{\frac{p+1}{8}\alpha + \frac{\alpha}{6-p}}
\left(dr^2 + r^2
d\Omega_{7-p}^2\right) + \left(\frac{H}{\tilde{H}}\right)^
{-\frac{7-p}{8}\alpha}(-dt^2 + \sum_{i=1}^{p+1}
dx_i^2)\nn
e^{2\phi} &=& 
\left(\frac{H}{\tilde {H}}\right)^{-a\alpha+2\delta_1},\qquad
F_{[q]}\,\, =\,\, 0\\
T_{00} &=& -T_{ii} \,\,=\,\, \frac{\Omega_{7-p}}{2\kappa_0^2} (6-p)
\omega^{6-p}\left[\frac{2\alpha(7-p)}{6-p}\right]
\eea
This is the supergravity configuration of non-BPS D$(p+1)$-brane discussed
by Sen \cite{asone} and were also obtained in refs.\cite{bmo,divec,lrone}. 
The solutions in this case
are characterized by two parameters $\omega$ and $\alpha$ (or $\delta_2$). 
The parameter relation (2.17) takes the form,
\be
\delta_1^2 + \alpha(\alpha-a\delta_1) = \frac{(4-\alpha^2)(7-p)}{2(6-p)}
\ee
This determines $\delta_1$ in terms of $\alpha$ (or $\delta_2$) and
to have real and negative $\delta_1$, we must have
\be
 |\alpha| \le \frac{8}{\sqrt{2(5p + 14 - p^2)}}.
\ee 
We also
point out that for $p$ = even (odd) the original delocalized $p$-branes
in (2.22) represent solutions in type IIA (IIB) string theory. But since
we made the solutions to localized $(p+1)$-branes without T-duality 
transformation then the solutions in (2.26) also represent solutions in
the same theory i.e. in type IIA (IIB) theory for $p$ = odd (even). This
clarifies the reason why the non-BPS branes (of the type discussed by Sen)
in type IIA and IIB string
theories have the wrong dimensionalities compared to the BPS branes.

Now in order to see how the delocalized solutions (2.22) reduce to BPS 
$p$-branes, we need the necessary condition
from the expressions of  the e-m tensors
in (2.25),  $T_{xx}=0$ and $T_{00}=-T_{ii}$ for $i=1,\ldots,p$. This
condition means we take either $\delta_2 \to 0$ or $\omega \to
0$. Examining the metric (2.22) carefully, we have the correct BPS limit
by sending 
$|\theta| \to  \infty$ while having $\omega \to 0$ according to the
following
\bea
 \omega^{6-p} & \to & \epsilon\,\bar{\omega}^{6-p}\nn
(\alpha + \beta) \sinh2\theta & \to & \epsilon^{-1}
\eea
where $\epsilon$ is a dimensionless parameter which tends to zero. With the
above scaling $b \to (6-p) \bar{\omega}^{6-p}$, $F \to \bar{H} = 1 +
\frac{\bar{\omega}^{6-p}}{r^{6-p}}$, and $H,\, \tilde{H} \to 1$. Since both
$\delta_1$ and $\delta_2$ are bounded given in (2.19), it can be easily
checked that the configuration (2.22) reduce to
\bea
ds^2 &=& \bar{H}^{\frac{p+1}{8}}
\left(dr^2 + r^2
d\Omega_{7-p}^2 + dx_{p+1}^2\right) + \bar{H}^
{-\frac{7-p}{8}}(-dt^2 + \sum_{i=1}^{p}
dx_i^2)\nn
e^{2\phi} &=& 
\bar{H}^{-a},\qquad
F_{[q]}\,\, =\,\, b {\rm Vol}(\Omega_{7-p}) \wedge dx_{p+1}
\eea
This is precisely the BPS D$p$-brane solutions delocalized in $x_{p+1}$ 
direction. Note from (2.25) that even in this case $T_{xx} \to 0$ and
$T_{00} = -T_{ii} \to \frac{\Omega_{7-p}}{2\kappa_0^2} (6-p) 
\bar{\omega}^{6-p}$. However, this delocalization is trivial in the sense that
since $T_{xx}=0$, we can always replace the line source along $x$-direction
by a delta function source without any cost of energy (true for BPS branes).
In other words, in calculating the e-m tensor we replace the Poisson's
equation of the harmonic function $\bar{H}$ as,
\be
\nabla^2\bar{H} = - \Omega_{7-p} (6-p) \bar{\omega}^{6-p}
\delta^{(8-p)}(r) \to 
\nabla^2\bar{H} = - \Omega_{8-p} (7-p) \bar{\omega}^{7-p}
\delta^{(9-p)}(r) 
\ee
The harmonic function now takes the form $\bar{H} = 1 + \bar{\omega}^{7-p}/
r^{7-p}$ where $r$ includes $x\equiv x_{p+1}$ The e-m tensor will be given as
$T_{00} = -T_{ii} \to \frac{\Omega_{8-p}}{2\kappa_0^2} (7-p) 
\bar{\omega}^{7-p}$, for $i=1,\ldots,p$ and (2.30) will reduce to the 
localized D$p$-brane solutions.

This therefore shows how the delocalized, non-supersymmetric $p$-brane 
solutions (2.22) can be regarded as the interpolating solutions between
the non-BPS D$(p+1)$-branes and BPS D$p$-branes very similar to the tachyon
condensation on the tachyonic kink solution on the non-BPS branes.

\sect{Wick rotation and delocalized, non-SUSY $p$-branes}

In this section we will Wick rotate the spatially delocalized solutions
(2.22) and obtain temporally delocalized $p$-branes as follows. Let us 
make the following Wick rotation,
\bea
x_{p+1} &\to& it\nn
t &\to& i x_{p+1}
\eea
Since the harmonic functions $H$ and $\tilde{H}$ are independent of both 
$x_{p+1}$ and $t$, under the above change the configurations (2.22) become,
\bea
ds^2 &=& F^{\frac{p+1}{8}} (H{\tilde {H}})^{\frac{2}{6-p}}\left(
\frac{H}{\tilde{H}}\right)^{-\frac{2\delta_2}{6-p}}
\left(dr^2 + r^2
d\Omega_{7-p}^2\right) + F^{-\frac{7-p}{8}} \sum_{i=1}^{p+1}
dx_i^2
%\nn
%& & \qquad\qquad\qquad\qquad\qquad\qquad\qquad\qquad\qquad + 
-F^{\frac{p+1}{8}}\left(
\frac{H}{\tilde{H}}\right)^{2\delta_2} dt^2\nn
e^{2\phi} &=& F^{-a}
\left(\frac{H}{\tilde {H}}\right)^{2\delta_1},\qquad
F_{[q]}\,\, =\,\, i b\,\,{\rm Vol}(\Omega_{7-p}) \wedge dt
\eea
Note that under the Wick rotation (3.1) the field strength has become 
imaginary and so, if we insist on real solutions $b$ must vanish or in other
words, the solutions in this case must be chargeless. $b=0$ implies 
$\theta=0$ by (2.16) and so, $F$ in (2.15) takes the form,
\be
F = \left(\frac{H}{\tilde{H}}\right)^\alpha
\ee
So, the real solutions in this case become,
\bea
ds^2 &=& (H{\tilde {H}})^{\frac{2}{6-p}}\left(
\frac{H}{\tilde{H}}\right)^{\frac{p+1}{8}\alpha-\frac{2\delta_2}{6-p}}
\left(dr^2 + r^2
d\Omega_{7-p}^2\right) + \left(\frac{H}{\tilde{H}}\right)^
{-\frac{7-p}{8}\alpha} \sum_{i=1}^{p+1}
dx_i^2
\nn
& & \qquad\qquad\qquad\qquad\qquad\qquad\qquad\qquad\qquad  
-\left(\frac{H}{\tilde{H}}\right)^{\frac{p+1}{8}\alpha + 2\delta_2}dt^2\nn
e^{2\phi} &=& 
\left(\frac{H}{\tilde {H}}\right)^{-a\alpha+2\delta_1},\qquad
F_{[q]}\,\, =\,\,0 
\eea
Therefore, unlike the case of spatially delocalized solutions, the temporally 
delocalized solutions are characterized by three parameters $\omega$, 
$\delta_1$ and $\delta_2$ ($\alpha$ is related to $\delta_1$ and $\delta_2$
by (2.18)). The brane directions in (3.4) are all spatial and so they are
Euclidean branes (or S-branes) delocalized in the transverse time-like
direction. However, because these solutions  
are chargeless, it is not possible
to obtain the localized S-branes \cite{stro,cgg,kmp,roy,ohta}
by localizing the time direction as was
done for the case of spatially delocalized solutions.

As before we calculate the various components of the e-m tensor from the 
metric in (3.4) as,
\bea
T_{00} &=& -\frac{\Omega_{7-p}}{2\kappa_0^2} (6-p)
\omega^{6-p}\left[\frac{4\delta_2 (7-p)}{6-p}\right]\nn
T_{ij} &=& -\frac{\Omega_{7-p}}{2\kappa_0^2} (6-p)
\omega^{6-p}\left[2\alpha -\frac{4\delta_2}{6-p}\right]\delta_{ij}\nn
T_{mm} &=& 0
\eea
where in the above $i,j=1,\ldots,p+1$ and $m=p+2,\ldots,9$. We also note
from (3.5) that for $\alpha=-2\delta_2$, $T_{00}=-T_{ii}$ as expected of a
localized $(p+1)$-brane. Indeed we find that under this condition, the 
coefficient of $-dt^2$ and the coefficient of $(dx_1^2+\cdots+dx_{p+1}^2)$
term match. The configurations (3.4) in this case reduce to
\bea
ds^2 &=& (H{\tilde {H}})^{\frac{2}{6-p}}\left(
\frac{H}{\tilde{H}}\right)^{\frac{p+1}{8}\alpha + \frac{\alpha}{6-p}}
\left(dr^2 + r^2
d\Omega_{7-p}^2\right) + \left(\frac{H}{\tilde{H}}\right)^
{-\frac{7-p}{8}\alpha}(-dt^2 + \sum_{i=1}^{p+1}
dx_i^2)\nn
e^{2\phi} &=& 
\left(\frac{H}{\tilde {H}}\right)^{-a\alpha+2\delta_1},\qquad
F_{[q]}\,\, =\,\, 0
\eea
This is precisely the non-BPS D$(p+1)$-brane solutions obtained before
in (2.26), although our starting solutions in these two cases are different.

On the other hand, we note that the time direction can not be made true
transverse direction of the brane by adjusting or scaling the parameters
as was done for the spatially delocalized solutions. From the e-m tensors
in (3.5), however, it might seem that it is possible to achieve that either
for $\delta_2 \to 0$ or for
$\omega^{6-p} \to 0$, when $T_{00}$ vanishes. (Note that this happens for 
S-branes 
where time is the true transverse direction of the Euclidean or S-branes.) 
But
it is clear from the metric in (3.4) that the coefficients of $-dt^2$ term
and $(dr^2 + r^2 d\Omega_{7-p}^2)$ term do not match for $\delta_2=0$. So,
even if $T_{00}$ vanishes, `$t$' does not become a transverse direction
of the brane. This happens because $T_{00}$ encodes
only the linear property of the metric. In the other limit $\omega^{6-p}
\to 0$, $T_{00} \to 0$, but since $\alpha$ and $\delta_2$ are finite, we
also have $T_{ij} \to 0$ and the metric becomes trivial i.e. the flat
space. Note that this did not happen for the spatially delocalized solutions
because the solutions were charged and involved more parameters which
were scaled appropriately to obtain the localized BPS D$p$-brane solutions.
However, in this case we can keep $T_{00}$ = fixed and send $T_{ij} \to 0$
by allowing $\alpha \to 2\delta_2/(6-p)$ (Note that this is possible
only when the minus sign is chosen in (2.18).). This is exactly the 
configuration one
gets for the tachyon dust or tachyon matter \cite{asfive}
which is pressureless and possesses
fixed energy. This configuration is possible because of the extra parameter
$\delta_2$ in the solutions. Now with the condition
\be
\alpha = \frac{2\delta_2}{6-p}
\ee
the solutions (3.4) will be characterized by two parameters only, namely,
$\omega$ and $\delta_1$ (since $\delta_1$ and $\delta_2$ are related by
(2.18)). However, we would like to point out that this is not the end of
the story. It is known from the string field theory \cite{st,jm}
as well as the tachyon 
effective action analysis \cite{assix}, that for the rolling 
tachyon, when the tachyon 
condenses, not only the pressure vanishes, but also the so-called dilaton 
charge vanishes \cite{oy}. The dilaton charge is the source for 
the dilaton equation
of motion following from the action (2.1). Since the value of the dilaton 
charge is frame dependent we work in the string frame where the metric is
$\hat{g}_{\mu\nu} = e^{\phi/2} g_{\mu\nu}$. Expanding the string frame
metric around the asymptotically flat region $\hat{g}_{\mu\nu} = \eta_{\mu\nu}
+ \hat{h}_{\mu\nu}$, we get from (3.4)
\bea
\hat{h}_{00} &=& \left[\alpha - \delta_1 - 2(7-p)\alpha\right]
\frac{\omega^{6-p}}{r^{6-p}}\nn 
\hat{h}_{ij} &=& (\delta_1 - \alpha) 
\frac{\omega^{6-p}}{r^{6-p}}\delta_{ij}, \qquad\qquad i,j = 1,\ldots,p+1\nn 
\hat{h}_{mn} &=& (\delta_1 - \alpha) 
\frac{\omega^{6-p}}{r^{6-p}}\delta_{mn}, \qquad\qquad m,n = p+2,\ldots,9\nn 
\hat{h} &=& \eta^{\mu\nu} \hat{h}_{\mu\nu}\,\,=\,\, 2\left[5\delta_1
-\alpha (p-2)\right] \frac{\omega^{6-p}}{r^{6-p}}
\eea
The linearized equation of motion of the dilaton in the string frame takes 
the form
\be
\nabla^2\left(\hat{h}_{mm} - \hat{h} + 4\phi\right) = -2\kappa_0^2 
Q_D \delta^{(8-p)}(r)
\ee
Whence we obtain,
\be
Q_D = \frac{\Omega_{7-p}}{2\kappa_0^2} (6-p) (\alpha-\delta_1)\omega^{6-p}
\ee
where $\phi$ was calculated from (3.4) as $\phi=(2\delta_1 - \frac{p-3}{2}
\alpha)\frac{\omega^{6-p}}{r^{6-p}}$. Thus we find that for the dilaton charge
to vanish
\be
\alpha = \delta_1
\ee
Using (3.7) and (3.11) we obtain from (2.18) in $d=10$
\be
\alpha = -  \sqrt{\frac{4}{(6-p)(7-p)}} = \delta_1 = \frac{2\delta_2}{6-p}
\ee
Substituting (3.12) into (3.4) we find the supergravity configuration of 
tachyon matter as, 
\bea
ds^2 &=& \left(\frac{H}{\tilde{H}}\right)^{\frac{1}{4}\sqrt{\frac{7-p}
{6-p}}}
\left[-\left(\frac{H}{\tilde{H}}\right)^{- 2 \sqrt{\frac{7-p}{6-p}}} dt^2
+ \sum_{i=1}^{p+1} (dx^i)^2 + (H\tilde{H})^{\frac{2}{6-p}}
\left(dr^2 + r^2 d\Omega_{7-p}^2\right)\right]\nn
e^{2\phi} &=& \left(\frac{H}{\tilde{H}}\right)^{- 
\sqrt{\frac{7-p}{6-p}}},\qquad F_{[8-p]}\,\,=\,\,0
\eea
This is precisely the tachyon matter configuration obtained in \cite{oy}
using a different method. We have thus seen how the Wick rotated solutions
or the temporally delocalized non-supersymmetric $p$-branes (3.4) can be 
regarded as the interpolating solutions between non-BPS D$(p+1)$-branes
and the tachyon matter, similar to the picture of rolling tachyon 
\cite{asfive} on the
non-BPS D-branes discussed by Sen. We like to point out that although the 
rolling tachyon implies that the tachyon is time dependent, the supergravity
solutions are still static. The reason is, the supergravity configurations
represent S-branes delocalized in the time direction and unless the
time direction is fully localized the supergravity configurations will 
remain static. In the approach of \cite{oy}, the question of time independence
of the supergravity configurations or even why one should start with
the non-supersymmetric black $p$-brane solutions \cite{zz} to arrive 
at tachyon matter
was not clear. Our approach, however, clarifies these points.

\sect{Conclusion}

To summarize, in this paper we have constructed non-supersymmetric spatially
delocalized (in one transverse direction) $p$-brane solutions of type II
supergravities in $d$ space-time dimensions. Unlike the localized solutions
(which contain three parameters), the delocalized solutions are 
characterized by four parameters. We have shown how these solutions in
$d=10$ nicely interpolate between non-BPS D$(p+1)$-branes (of the type
discussed by Sen) and the BPS D$p$-branes. This process is very similar
to the picture of tachyon condensation on the tachyonic kink solution of the
non-BPS D$(p+1)$-branes. In our approach we have clarified the reasons
for the appearance of even and odd dimensional non-BPS D-branes in type
IIB and type IIA string theories respectively. Further, we have obtained
non-supersymmetric, temporally delocalized Euclidean $p$-brane solutions
by an Wick rotation on our previous solutions. We have shown how these latter
solutions nicely interpolate between non-BPS D$(p+1)$-branes and the 
tachyon matter supergravity configurations. This process is very similar
to the picture of rolling tachyon on the non-BPS D$(p+1)$-branes. Our approach
also clarifies why we need a static solution to understand the tachyon 
condensation for the time-dependent tachyon or the rolling tachyon on the
non-BPS D-branes. We emphasize that although we have indicated the 
similarities of our approach with the process of tachyon condensation for
the space dependent as well as the time dependent tachyons, it would be nice
to understand the physical meanings of the parameters and the exact 
relationships of them with the dynamics of tachyon condensation.
\vskip .5cm

\noindent
{\bf Acknowledgements}

    JXL would like to thank the Michigan Center for Theoretical Physics
 for hospitality and partial support during the final stage of this
 work. He also acknowledges support by grants from the Chinese Academy
 of Sciences and the grant from the NSF of China with Grant No: 90303002.

\end{document}